\documentclass[showpacs,preprintnumbers,amsmath,amssymb]{revtex4}
\usepackage{amsthm}
\usepackage{enumerate}
\usepackage{epsfig}
 \newtheorem{theorem}{Theorem}[section]
 
 \newtheorem{corollary}[theorem]{Corollary}
 \newtheorem{definition}[theorem]{Definition}
 \newtheorem{proposition}[theorem]{Proposition}

\newtheorem{example}[theorem]{Example}

\newcommand*{\complex}{\bbC}

\newcommand*{\cD}{\mathcal{D}}

\newcommand*{\cF}{\mathcal{F}}

\newcommand*{\cI}{\mathcal{I}}

\newcommand*{\cU}{\mathcal{U}}

\newcommand*{\symtwirl}{\mathbb{S}}
\newcommand*{\twirl}{\mathbb{T}}

\newcommand*{\tr}{\mathsf{tr}}

\newcommand*{\ket}[1]{|#1\rangle}
\newcommand*{\bra}[1]{\langle #1|}
\newcommand*{\proj}[1]{\ket{#1}\bra{#1}}

\newcommand*{\spec}[1]{\mathsf{Spec(#1)}}

\newcommand*{\Par}[2]{Par(#1,#2)}

\newcommand{\braket}[2]{\langle #1|#2\rangle}       % Inner Product

\newcommand{\be}{\begin{equation}}
\newcommand{\ee}{\end{equation}}
\newcommand{\bea}{\begin{eqnarray}}
\newcommand{\eea}{\end{eqnarray}}
\newcommand{\bestar}{\begin{equation*}}
\newcommand{\eestar}{\end{equation*}}
\newcommand{\beastar}{\begin{eqnarray*}}
\newcommand{\eeastar}{\end{eqnarray*}}

\def\complex{\mathbb{C}}

\begin{document}

\title{A dual de Finetti theorem}

\author{Graeme \surname{Mitchison}}
\email[]{g.j.mitchison@damtp.cam.ac.uk} \affiliation{Centre for
Quantum Computation, DAMTP,
             University of Cambridge,
             Cambridge CB3 0WA, UK}

\begin{abstract}
The quantum de Finetti theorem says that, given a symmetric state, the
state obtained by tracing out some of its subsystems approximates a
convex sum of power states. The more subsystems are traced out, the
better this approximation becomes. Schur-Weyl duality suggests that
there ought to be a dual result that applies to a unitarily invariant
state rather than a symmetric state. Instead of tracing out a number
of subsystems, one traces out part of every subsystem. The theorem
then asserts that the resulting state approximates the fully mixed
state, and the larger the dimension of the traced-out part of each
subsystem, the better this approximation becomes. This paper gives a
number of propositions together with their dual versions, to show how
far the duality holds.
\end{abstract}

\pacs{03.67.-a, 02.20.Qs}

\date{\today}

\maketitle

\pagestyle{plain}

\section{Introduction}

Suppose we have a state space $H=(\complex^d)^{\otimes n}$ consisting
of $n$ identical subsystems. The quantum de Finetti theorem
~\cite{KoeRen05,Ren05} tells us that, given a symmetric state on $H$,
the state obtained by tracing out $n-k$ of the subsystems can be
approximated by a convex sum of powers, i.e. by a convex sum of states
of the form $\sigma^{\otimes k}$; the smaller $k/n$, the better the
approximation. This is a useful result, because such power states are
often rather easy to analyse.

Now, the symmetric group $S_n$ and the unitary group ${\cal U}(d)$
both act on the space $(\complex^d)^{\otimes n}$, the former by
permuting the factors and the latter by applying any $g \in {\cal
U}(d)$ to each factor, so the action is given by $g^{\otimes
n}$. These actions commute, and this leads to a type of duality,
called Schur-Weyl duality \cite{GooWal98}. Given any result that
holds for the symmetric group, one can hope to find a dual result for
the unitary group.

Here I show that there is a dual to the de Finetti theorem, obtained
by swapping the roles of $S_n$ and ${\cal U}(d)$.  The situation is
summed up in Table \ref{dual1}. Instead of symmetric states, we
consider unitarily-invariant states. And instead of tracing out a
number of subsystems, we trace out part of each subsystem; more
precisely, we replace each individual subsystem $\complex^d$ by
$\complex^p \otimes \complex^q$, and we trace out the $\complex^q$
part from all the subsystems in $(\complex^p \otimes
\complex^q)^{\otimes n}$. The theorem then states that, when $q$ is
large relative to $n$, the resulting traced-out state approximates the
fully mixed state. This is different in character from the standard de
Finetti theorem, in that all information about the original state is
lost. However, this fact in itself may lead to some interesting
applications.

As far as possible, the results are laid out as pairs of propositions
that are duals of each other. Some of these pairs are exact analogues;
in other cases, one of the pair is less meaningful or even
trivial. This gives some insight into the nature of the duality.

\begin{table}
\caption{
\label{dual1}}
\begin{center}
\begin{tabular}{|c|c|}
\hline
{\bf Standard de Finetti theorem.} & {\bf Dual theorem.}\\
\hline
Symmetric state $\rho$. & Unitarily-invariant state $\rho$.\\
\hline
State space is $(\complex^d)^{\otimes n}$. &  State space is $(\complex^p \otimes \complex^q)^{\otimes n}$.\\
\hline
Trace out $n-k$ subsystems. & Trace out $\complex^q$ from each subsystem.\\
\hline
 $\tr_{n-k} \rho \approx \mbox{convex sum of powers.}$ &   $\tr_{\complex^q} \rho \approx \mbox{fully mixed state.}$\\
\hline
\end{tabular}
\end{center}
\end{table}
%
%
%
%               SYMMETRIC WERNER STATES
%
%
%
\section{Duality for symmetric Werner states.}

We will refer to unitarily-invariant states as {\em Werner} states
\cite{Werner89}. Rather than considering general Werner states, we
begin by looking at a special class, the symmetric Werner states,
i.e. states that are invariant under both the unitary and symmetric
groups. The de Finetti theorem and its dual can then be applied to the
same state, so the pattern becomes particular clear, as shown in Table
\ref{dual2}.

The Schur-Weyl decomposition \cite{FultonHarris91} of
$H=(\complex^d)^{\otimes n}$ is given by:
\be \label{SchurWeyl} 
(\complex^d)^{\otimes n} \cong
\bigoplus_{\lambda\in \Par{n}{d}} U_\lambda \otimes V_\lambda,
\ee
where $U_\lambda$ is the irrep (irreducible representation) of
$\cU(d)$ with highest weight $\lambda_1 \ge \lambda_2 \ge \ldots \ge
\lambda_d$, and $V_\lambda$ is the irrep of $S_n$ defined by the same
partition $\lambda$. Here $\Par{n}{d}$ denotes the ordered partitions
of $n$ with at most $d$ rows. We will also refer to a $\lambda \in
\Par{n}{d}$ as a (Young) diagram.

Let $P_\lambda$ denote the projector onto the subspace $U_\lambda
\otimes V_\lambda$ in the Schur-Weyl decomposition. Write
$f_\lambda=\dim(V_\lambda)$, and $e^d_\lambda=\dim(U_\lambda)$, where
$d$ is the dimension of the unitary group $\cU(d)$. Then the
normalised projector $\rho_\lambda=P_\lambda/(e^d_\lambda f_\lambda)$
is a symmetric Werner state, and in fact any symmetric Werner state
$\rho$ can be written as a weighted sum of such projectors, $\sum_\mu
a_\mu \rho_\mu$, with $\sum_\mu a_\mu=1$ \cite{ChrEtal06}. Let
$\tr_{n-k} \rho_\lambda$ denote the state obtained by tracing out
$n-k$ of the $n$ subsystems from the state $\rho_\lambda$. Lemma III.4
in \cite{ChrEtal06} can be restated as follows:
\begin{proposition}[Trace formula] \label{original-trace}
Let $\lambda\in\Par{n}{d}$. Then
\[
\tr_{n-k} \rho_\lambda =\frac{1}{f_\lambda} \sum_\mu \rho_\mu f_\mu
\left(\sum_\nu c_{\mu \nu}^\lambda f_\nu\right),
\]
 where the sums extends over all $\mu \in \Par{k}{d}$ and $\nu \in
 \Par{n-k}{d}$, and $c_{\mu \nu}^\lambda$ is the Littlewood-Richardson
 coefficient, i.e. the coefficient in the Clebsch-Gordan series for
 $\cU(d)$: $U_\mu \otimes U_\nu=\sum_\lambda c^\lambda_{\mu\nu}
 U_\lambda$.
\end{proposition}

From now on, we assume each individual subsystem $\complex^d$ is
bipartite, so it can be written as $\complex^p \otimes
\complex^q$. Let $\tr_{\complex^q}$ denote the result of tracing out
$\complex^q$ from each subsystem in the total state space $(\complex^p
\otimes \complex^q)^{\otimes n}$. The dual of the preceding
Proposition is:
\begin{proposition}[Dual trace formula] \label{dual-trace}
Let $\lambda\in\Par{n}{pq}$. Then
\[
\tr_{\complex^q} \rho_\lambda =\frac{1}{e^{pq}_\lambda} \sum_\mu \rho_\mu e^p_\mu \left(\sum_\nu g_{\lambda \mu \nu} e^q_\nu
\right),
\]
where the sums extend over all diagrams $\mu \in \Par{n}{p}$ and $\nu
\in \Par{n}{q}$, and $g_{\lambda \mu \nu}$ is the Kronecker
coefficient, i.e. the coefficient in the Clebsch-Gordan series for
$S_n$: $V_\mu \otimes V_\nu=\sum_\lambda g_{\lambda\mu\nu}
V_\lambda$.
\end{proposition}
\begin{proof}
We can restrict the action of the group $\cU(pq)$ on $\complex^p
\otimes \complex^q$ to the subgroup $\cU(p) \times \cU(q)$. This
gives an expansion in tensor products of irreps \cite{ChrMit06}:
\[
U_\lambda = \sum_{\mu \nu} g_{\lambda \mu \nu} U_\mu \otimes U_\nu,
\]
where $\mu \in \Par{n}{p}$ and $\nu \in \Par{n}{q}$.  If
$P_{U_\lambda}$ denotes the projector onto $U_\lambda$, we can rewrite
this as
\be \label{content-projection} 
P_{U_\lambda} = \sum_{\mu \nu} \sum_{i=1}^{g_{\lambda \mu \nu}} P^i_{U_\mu} \otimes P^i_{U_\nu}.
\ee 
Taking the trace over $\complex^q$ gives
\be \label{content-trace}
\tr_{\complex^q} P_{U_\lambda} = \sum_{\mu \nu} \sum_{i=1}^{g_{\lambda \mu \nu}} P^i_{U_\mu}  e^q_\nu.
\ee 
Now define the symmetric average, $\symtwirl$, by
\be \label{stwirl}
\symtwirl(\tau)=\frac{1}{n!}\sum_{\pi \in S_n}\pi \tau \pi^{-1},
\ee
for any operator $\tau$. Applying $\symtwirl$ to both sides of
(\ref{content-trace}), Schur's lemma implies
\[
\tr_{\complex^q}
\frac{P_\lambda}{f_\lambda} = \sum_{\mu}
\frac{P_\mu}{f_\mu}\left(\sum_\nu g_{\lambda \mu \nu} e^q_\nu \right).
\]
Substituting $\rho_\lambda=P_\lambda/(e^{pq}_\lambda f_\lambda)$,
$\rho_\mu=P_\mu/(e^p_\mu f_\mu)$ gives the result we seek.
\end{proof}
This shows, incidentally, why the dual operation to tracing out over
$n-k$ subsystems is to trace out over part of each subsystem: the
analogue of the subgroup $S_k \times S_{n-k} \subset S_n$ is the
subgroup $\cU(p) \times \cU(q) \subset \cU(pq)$.

\begin{table}
\caption{
\label{dual2}}
\begin{center}
\begin{tabular}{|c|c|}
\hline
\multicolumn{2}{|c|}{{\bf Duality Dictionary for symmetric Werner states.}}\\
\hline
$f_\lambda \ \ (\dim V_\lambda).$ & $e^d_\lambda \ \ (\dim U_\lambda).$ \\
\hline
Littlewood-Richardson coefficient $c^\lambda_{\mu \nu}$. & Kronecker coefficient $g_{\lambda \mu \nu}$.\\
\hline
Unitary group character (Schur function) $s_\lambda$. &  Symmetric group character $\chi^\lambda$.\\
\hline
Shifted Schur function $s^*_\mu(\lambda)$. &   Character polynomial $\chi^{\lambda \mu}(q)$ (Definition \ref{character-polynomial-definition}).\\
\hline
Twirled power state. &  Symmetrised cycle operator.\\
\hline
\end{tabular}
\end{center}
\end{table}

Theorem 8.1 in \cite{OkoOls96} allows one to evaluate the bracketted
inner sum in Proposition \ref{original-trace}. We restate this result
as follows:
\begin{proposition}[Inner sum formula] \label{fsum}
\[
\sum_\nu c_{\mu \nu}^\lambda f_\nu = \frac{f_\lambda
s_\mu^*(\lambda)}{n \downharpoonright k},
\]
where $s_\mu^*(\lambda)$ is the shifted Schur function defined in
\cite{OkoOls96} and $n \downharpoonright k =n(n-1) \ldots (n-k+1)$.
\end{proposition}
Likewise, one can evaluate the bracketed inner sum in Proposition
\ref{dual-trace}. First we introduce a symmetric-group analogue of the
shifted Schur function:
\begin{definition} \label{character-polynomial-definition}
Suppose $\lambda$ and $\mu$ are arbitrary diagrams with $n$ boxes. The
{\em character polynomial} $\chi^{\lambda \mu}(q)$ is the polynomial
in $q$ defined by
\[
\chi^{\lambda \mu}(q)=\sum_{\pi \in S_n}
q^{c(\pi)} \chi^\lambda(\pi) \chi^\mu(\pi),
\]
where $\chi^\mu(\pi)$ is the character of the symmetric group evaluated
at the permutation $\pi$ and  $c(\pi)$ is the number of cycles in $\pi$.
\end{definition}
The character polynomial can sometimes be more conveniently calculated by
summing over cycle types $\alpha$ rather than permutations, giving
\[ 
 \chi^{\lambda \mu}(q)=\sum_{\alpha \in \Par{n}{n}} h_\alpha q^{c(\alpha)} \chi^\lambda(\alpha)
\chi^\mu(\alpha), 
\]
where $h_\alpha$ is the number of elements in the conjugacy class
$\alpha$ \cite{digest}, and $c(\alpha)$ is the number of rows in the
diagram $\alpha$ representing the cycle type.
\begin{proposition}[Dual inner sum formula] \label{gsum}
\[
\sum_\nu g_{\lambda \mu \nu} e^q_\nu= \frac{\chi^{\lambda \mu}(q)}{n!}.
\]
\end{proposition}
\begin{proof}
First observe that 
\be \label{e-formula}
e^q_\nu=\frac{1}{n!} \sum_{\pi \in S_n}
q^{c(\pi)} \chi^\nu(\pi).
\ee
This follows from the fact \cite{digest} that the projector $P_\nu$ on
$(\complex^q)^{\otimes n}$ is defined by
\[
P_\nu=\frac{f_\nu}{n!}\sum_\pi \chi^\nu(\pi) \pi,
\]
and it vanishes on all components of the Schur-Weyl decomposition
(\ref{SchurWeyl}) except $U_\nu \otimes V_\nu$, where it has trace
$e^q_\nu f_\nu$. On the other hand, the trace of $\pi$ acting on
$(\complex^q)^{\otimes n}$ is given by $q^{c(\pi)}$ since the basis
elements $e_{i_1} \otimes \ldots \otimes e_{i_n}$ of
$(\complex^q)^{\otimes n}$ that are fixed by $\pi$, i.e. that
contribute to $\tr \pi$, are those that assign the same $e_i$ to all
the elements of each cycle of $\pi$, and there are $q$ ways of picking
an $e_i$ and $c(\pi)$ cycles. Thus $P_\nu$ has trace $\frac{f_\nu}{n!}
\sum_\pi q^{c(\pi)}\chi^\nu(\pi)$, and equating these two expressions
for the trace gives (\ref{e-formula}).

Now the Kronecker coefficient can be defined by
\[
g_{\lambda \mu \nu}=\frac{1}{n!}\sum_\pi \chi^\lambda(\pi)\chi^\mu(\pi)\chi^\nu(\pi).
\]
Combining this with (\ref{e-formula}), we have
\[
\sum_\nu g_{\lambda \mu \nu} e^q_\nu= \frac{1}{n!} \sum_{\pi, \pi^\prime} 
q^{c(\pi^\prime)} \chi^\lambda(\pi) \chi^\mu(\pi)\left(\frac{1}{n!} \sum_\nu
\chi^\nu(\pi) \chi^\nu(\pi^\prime) \right).
\]
The orthogonality relations for characters imply that the expression
in brackets is zero if $\pi$ and $\pi^\prime$ are in different
conjugacy classes, and is otherwise the inverse of $h_{[\pi]}$, the
number of elements in the conjugacy class of $\pi$. As $c(\pi^\prime)$
only depends on the conjugacy class of $\pi^\prime$, the result
follows.
\end{proof}

Propositions \ref{fsum} and \ref{original-trace} can be
used to prove the de Finetti theorem for symmetric Werner states
\cite{ChrEtal06}:
\begin{theorem}[de Finetti theorem] \label{deFinetti}
Let $\rho_\lambda$ be the normalised projector onto the Young subspace
of $(\complex^d)^{\otimes n}$ with diagram $\lambda$. Then
\[
||\tr_{n-k} \rho_\lambda-\tau|| \leq\frac{3}{4}\cdot
\frac{k(k-1)}{\lambda_\ell}+O(\frac{k^4}{\lambda_\ell^2})\ ,
\]
where $\tau$ is a convex sum of power states and $\lambda_\ell$ is the
smallest non-zero component of $\lambda$.
\end{theorem}

\begin{theorem}[Dual de Finetti theorem] \label{dual-deFinetti}
\[
||\tr_{C^q}\rho_\lambda - \frac{\cI}{p^n}|| \le  2-2\left(\frac{q-n+1}{q}\right)^n = \frac{2n(n-1)}{q} + O(n^4/q^2),
\]
where $\cI$ is the identity on $(\complex^p)^{\otimes n}$.
\end{theorem}
We leave the proof till section \ref{general}, where the theorem is
proved for all Werner states, not just symmetric ones.

\begin{example}
The simplest example is the symmetric subspace for $n=2$. Using
Propostions \ref{dual-trace} and \ref{gsum}, we find
\[
\tr_{\complex^q} \rho_{(2)}=\frac{(p+1)(q+1)}{2(pq+1)} \rho_{(2)}+\frac{(p-1)(q-1) }{2(pq+1)}\rho_{(1^2)}.
\]
Also 
\[
\frac{\cI}{p^2}=\frac{(p+1)}{2p}\rho_{(2)}+\frac{(p-1)}{2p} \rho_{(1^2)},
\]
from which one gets
\[
|| \tr_{\complex^q} \rho_{(2)}-\frac{\cI}{p^2}||=\frac{p^2-1}{p^2q+p}.
\]
Note that the bound tends to zero with $q \to \infty$ but its
behaviour does not depend sensitively upon $p$; in particular, there
is no requirement for $p$ to be small relative to $q$ (see
Discussion).
\end{example}

%
%
%
%                       TWIRLED POWER STATES
%
%
%

\section{Twirled power states and their duals}

Theorem \ref{deFinetti} in \cite{ChrEtal06} actually makes the
stronger claim that the approximating state $\tau$ is the twirl of a
power state $\sigma^k$. We describe this now and also its dual
version, where the analogue of the power state is a permutation
matrix. However, the rewards of the dual approach diminish rapidly,
and one does not get a stronger version of Theorem
\ref{dual-deFinetti} as will become clear at the end of this section.

Let us define the twirl of an arbitrary state $\tau$ on
$(\complex^d)^{\otimes k}$ as follows:
\[
\twirl(\tau)=\int U^{\otimes k} \tau (U^\dagger)^{\otimes k} dU,
\]
where $dU$ is the Harr measure. Suppose $r=(r_1, \ldots, r_d)$ is the
spectrum of a state $\sigma$ on $\complex^d$. Then the {\em twirled
power state} $\tau(r)=\twirl(\sigma^{\otimes k})$ depends only on $r$
and not on the particular state $\sigma$ chosen. Lemma III.1 from
\cite{ChrEtal06} expresses $\tau(r)$ in terms of basic Werner states.
\begin{proposition}[Twirl sum]\label{original-twirl}
Given a spectrum $r=(r_1, \ldots, r_d)$,
\[
\tau(r)=\sum_\mu f_\mu s_\mu(r) \rho_\mu,
\]
where $s_\mu(r)$ is the Schur function.
\end{proposition}
To define the dual version of a twirled power state, let $\pi$
be a permutation, and let $b_1, \ldots b_d$ be a basis in
$\complex^d$. Define the permutation matrix $\tau_\pi$ by
$\tau_\pi=\pi \cI$, i.e.
\[
\tau_\pi=\sum_{0 \le i_1 \ldots i_n \le d} \ket{(b_{i_{\pi(1)}}, \ldots, b_{i_{\pi(n)}})}\bra{(b_{i_1}, \ldots, b_{i_n})}.
\]
Let $\lambda$ be a Young diagram with $n$ boxes representing a
permutation cycle type. Pick any permutation $\pi$ with cycle type
$\lambda$. The {\em symmetrised cycle operator} $\sigma(\lambda)$ is
defined to be $\symtwirl(\tau_\pi/d^n)$, where $\symtwirl$ is defined by
(\ref{stwirl}). This does not depend on the choice of a permutation
$\pi$ having the cycle type $\lambda$. We can regard $\symtwirl$ as
the ``dual twirl'', with the symmetric group replacing the unitary
group. Thus we have:
\begin{proposition}[Dual twirl sum] \label{dual-twirl}
Given a cycle type $\lambda$, 
\be \label{dual-twirlsum}
\sigma(\lambda)=\frac{1}{d^n}\sum_\mu e^d_\mu \chi^\mu(\lambda) \rho_\mu.
\ee
\end{proposition}
\begin{proof}
By construction, $\sigma(\lambda)$ is symmetric; it is also unitarily
invariant, since $U\tau_\pi U^\dagger=U\pi I U^\dagger=\pi U I
U^\dagger=\pi I=\tau_\pi$. Thus $\sigma(\lambda)$ can be expressed as
a sum $\sum_\mu c_\mu \rho_\mu$, where the coefficients are given by
\[
c_\mu=\tr[P_\mu\sigma(\lambda)]=\tr[P_\mu\symtwirl(\tau_\pi/d^n)]=\tr[\symtwirl(P_\mu)\tau_\pi]/p^n=\tr[P_\mu \tau_\pi]/p^n,
\]
$\pi$ being a permutation of cycle type $\lambda$. But $\tr[P_\mu
\tau_\pi]$ is the character of the representation $\pi \to P_\mu
\tau_\pi P_\mu$, and as this is equivalent to $e^d_\mu$ copies of the
irrep $V_\lambda$, we have $c_\lambda=e^d_\mu\chi^\mu(\lambda)/p^n$.
\end{proof}
Note that $\sigma(\lambda)$ is in general not a state, since its
eigenvalues, the coefficients in (\ref{dual-twirlsum}), can be
negative. For instance, with $d=3$,
$\sigma((2,1))=\frac{10}{27}\rho_{(3)}-\frac{8}{27}\rho_{(1^3)}$.

Returning to the standard de Finetti theorem, Propositions
\ref{original-trace} and \ref{fsum} tell us that, for
$\lambda\in\Par{n}{d}$,
\[
\tr_{n-k} \rho_\lambda =\sum_\mu \rho_\mu f_\mu
\frac{s_\mu^*(\lambda)}{n \downharpoonright k},
\]
The shifted Schur function \cite{OkoOls96}, $s_\mu^*(\lambda)$, which
appears on the right-hand side of this equation, is a polynomial in
the $\lambda_i$, and its highest degree terms are the ordinary Schur
function $s_\mu(\lambda)$. It follows that 
\be \label{approx} 
\frac{s_\mu^*(\lambda)}{n
\downharpoonright k} \to s_\mu(\bar \lambda) \mbox{ as } n \to \infty, 
\ee
where $\bar \lambda=(\lambda_1/\sum \lambda_i, \ldots , \lambda_d/\sum
\lambda_i)$. Putting this together with Proposition
\ref{original-twirl}, we can restate Theorem \ref{deFinetti}, showing
that the approximating state can be taken to be the twirled power
state $\tau(\bar \lambda)$.
\begin{proposition}[Twirl limit for de Finetti theorem] \label{to-twirl} 
\be
||\tr_{n-k} \rho_\lambda - \tau(\bar \lambda)||\leq\frac{3}{4}\cdot
\frac{k(k-1)}{\lambda_\ell}+O(\frac{k^4}{\lambda_\ell^2})\ . 
\ee
\end{proposition}
Dually, Propositions \ref{dual-trace} and \ref{gsum} tell us that
\be \label{full-expansion}
\tr_{\complex^q} \rho_\lambda =\frac{1}{e^{pq}_\lambda} \sum_\mu \rho_\mu e^p_\mu \frac{\chi^{\lambda \mu}(q)}{n!},
\ee
We can imitate the approximation of the shifted Schur function by the
ordinary Schur function, and take the highest degree term in
$\chi^{\lambda \mu}(q)$, which is $q^n\chi^\lambda(1^n)\chi^\mu(1^n)$.
Using equation \ref{dual-twirlsum} and the fact that
$\chi^\lambda(1^n)=f_\lambda$, we get
\[
\tr_{\complex^q} \rho_\lambda \to \frac{(pq)^n f_\lambda}{e^{pq}_\lambda n!} \sigma(1^n) \mbox{ as } q \to \infty.
\]
We shall see later that the rather complicated coefficient of
$\sigma(1^n)$ tends to 1 for large $q$ (see inequality
\ref{finalbound}). This enables us to write
\begin{proposition}[Twirl limit for dual de Finetti theorem]
\be
||\tr_{C^q}\rho_\lambda - \sigma(1^n)|| \le   \frac{2n(n-1)}{q} + O(n^4/q^2).
\ee
\end{proposition}
Unlike Proposition \ref{to-twirl}, however, this adds nothing to the
preceding result (Theorem \ref{dual-deFinetti}), since
$\sigma(1^n)=\cI/p^n$. A more interesting result is obtained from
equation (\ref{full-expansion}) without taking the limit of
large $q$:
\be \label{cyclesum}
\tr_{\complex^q} \rho_\lambda=\frac{1}{n! e^{pq}_\lambda} \sum_\pi
 q^{c(\pi)} \chi^\lambda(\pi) \sigma(\pi),
\ee
This shows how symmetrised cycle operators other than $\sigma(1^n)$
contribute to the trace.
%
%
%
%               THE QUANTUM MARGINAL PROBLEM
%
%
%
\section{The quantum marginal problem and Horn's conjecture}

We have now compared most of the ingredients of the de Finetti theorem and
their dual versions. In this section we complete this process by
comparing the shifted Schur functions that appear in Proposition
\ref{fsum} with the character polynomials that appear in Proposition
\ref{gsum}. We do this by relating each of them to a mathematical
problem of some historical interest. For the shifted Schur functions
this is Horn's conjecture \cite{Knutson00}, whereas for the character
polynomials it is the quantum marginal problem \cite{Kly04}. We begin
with the latter. 

Let $\rho_A=\tr_B (\rho_{AB})$ and $\rho_B=\tr_A (\rho_{AB})$ be the
two marginal states of a bipartite state $\rho_{AB}$. Let
$\Sigma^{p,q}$ denote the set of triples of spectra
$\{\spec{\rho_{AB}}$, $\spec{\rho_A}$, $\spec{\rho_B}\}$ for all
operators $\rho_{AB}$ on $\complex^p \otimes \complex^q$. It was shown
in \cite{ChrMit06}, \cite{Kly04}, \cite{CHM06} that $\Sigma^{p,q}$ can
be defined in terms of the Kronecker coefficients. Given a diagram
$\lambda$, define $\bar \lambda=(\lambda_1/\sum \lambda_i, \ldots
\lambda_d/\sum \lambda_i)$, and let $K$ be the set of all triples
$(\bar \lambda, \bar \mu, \bar \nu)$ with $\lambda \in \Par{n}{pq}$,
$\mu \in \Par{n}{p}$, $\nu \in \Par{n}{q}$, for some $n$, satisfying
$g_{\lambda \mu \nu} >0$. Then $\Sigma^{p,q}$ is $\bar K$, the closure
of $K$.

One can also focus on a single marginal, and ask which {\em pairs},
$\{\spec{\rho_{AB}}$, $\spec{\rho_A}\}$ of spectra can occur
\cite{DH04}. From the characterisation of $\Sigma^{p,q}$, it follows
that this set, $\Gamma^{p,q}$ say, is the closure of the set of pairs
$(\bar \lambda, \bar \mu)$ where $\lambda \in \Par{n}{pq}$, $\mu \in
\Par{n}{p}$, and there is some $\nu \in \Par{n}{q}$ satisfying
$g_{\lambda \mu \nu} >0$. For a given $\lambda$, the $\mu$'s
satisfying this condition correspond to the $\rho_\mu$'s that have
non-zero coefficients in the expansion of $\tr_{\complex^q}
\rho_\lambda$ given by Proposition \ref{dual-trace}. This, together
with Proposition \ref{gsum}, implies
\begin{proposition}[Character polynomial condition for the marginal problem]\label{gamma}
Suppose $\lambda \in \Par{n}{pq}$, $\mu \in \Par{n}{p}$, and
$\chi^{\lambda \mu}(q) > 0$. Then $(\bar \lambda, \bar \mu) \in
\Gamma^{p,q}$.
\end{proposition}

The converse does not follow from the characterisation of
$\Sigma^{p,q}$ by Kronecker coefficients. If $\lambda \in \Par{n}{pq}$
and $\mu \in \Par{n}{p}$, and $(\bar \lambda, \bar \mu) \in
\Gamma^{p,q}$ then we know there is a state $\rho_{AB}$ with
$\spec{\rho_{AB}}=\bar \lambda$ and $\spec{\rho_A}=\bar \mu$, but it
does not follow that $\spec{\rho_B}$ has the form $\bar \nu$ for some
$\nu \in \Par{n}{q}$, or even that $\spec{\rho_B}$ is rational. Even
if it were true that $\spec{\rho_B}=\bar \nu$ with $\nu \in
\Par{n}{q}$, we could only conclude \cite{Kly04,CHM06} that 
$g_{m\lambda \ m\mu \ m\nu}>0$ for some integer $m>0$ and hence that
$\chi^{m\lambda \ m\mu}(q) > 0$ for some $m>0$.

%There are simple examples \cite{Kly04,Kirillov04} that
%show that $g_{m\lambda \ m\mu \ m\nu}>0$ does not imply $g_{\lambda
%\mu \nu}>0$; thus {\em saturation} (which holds for the
%Littlewood-Richardson coefficients \cite{KnuTao99}) fails for the
%individual Kronecker coefficients. However, it is possible that
%saturation holds for the sum $\sum_\nu g_{\lambda \mu \nu} e^q_\nu$.

\begin{proposition}  \label{zeros}
For any $\lambda \in \Par{n}{pq}$, $\mu \in \Par{n}{p}$, there is an
integer $q_+$ in the range $1 \le q_+ \le n$ such that
$\chi^{\lambda \mu}(q)>0$ for $q \ge q_+$ and $\chi^{\lambda
\mu}(q)=0$ for $0 \le q < q_+$. If $\lambda \ne \mu$, $q_+ \ge 2$.
\end{proposition}
\begin{proof}
Clearly $\chi^{\lambda \mu}(q)=0$ for $q=0$, and as
$\chi^{\lambda \mu}(q)$ is a polynomial of degree $n$ and can
therefore have at most $n$ distinct roots, there must be some integer
$q$ in the range $1 \le q \le n$ for which $\chi^{\lambda
\mu}(q)=0$. Let $q_+$ be the least such $q$. Then by Proposition
\ref{gsum}, $\sum_\nu g_{\lambda \mu \nu} e^{q_+}_\nu >0$, and thus
$g_{\lambda \mu \nu}>0$ and $e^{q_+}_\nu >0$ for some $\nu$. Thus
$e^q_\nu>0$ for all $q \ge q_+$, and $\chi^{\lambda \mu}(q)=\sum_\nu
g_{\lambda \mu \nu} e^q_\nu >0$ for all $q \ge q_+$.  If $\lambda \ne
\mu$, $\chi^{\lambda \mu}(1)=0$ by the orthogonality relations for
characters, so $q_+ \ge 2$.
\end{proof}
This result is also a consequence of a theorem of Berele and
Imbo \cite{BerImbo01}, which says that $g_{\lambda \mu \nu}>0$ for some
$\nu$ with $c(\nu) \le \max \{c(\lambda),c(\mu)\}$. This implies
the stronger result that $q_+ \le \max \{c(\lambda),c(\mu)\}$.

\begin{corollary} \label{nbound}
For any $\lambda \in \Par{n}{pq}$, $\mu \in \Par{n}{p}$, there is an
integer $q_+$ in the range $1 \le q_+ \le n$ such that $(\bar \lambda,
\bar \mu) \in \Gamma^{p,q}$ for $q \ge q_+$.
\end{corollary} 

\begin{example} \label{same}
Take $\lambda =\mu$. Since every term in $\chi^{\lambda \lambda}(1)$ is
non-negative, and the term with $\alpha=(1^n)$ is $f^2_\lambda/n!>0$,
we have $\chi^{\lambda \lambda}(1)>0$ and hence $(\bar \lambda, \bar
\lambda) \in \Gamma^{p,1}$. It is easy to see why this is true: take
$\rho_{AB}=\rho_A \otimes \proj{0}_B$, and
$\spec{\rho_{AB}}=\spec{\rho_A}$.
\end{example}

\begin{example} \label{extreme}
Take $\lambda=(1^n)$, $\mu=(n)$. Then
$\chi^\lambda(\pi)=(-1)^{n+c(\pi)}$, by the Murnaghan-Nakayama rule
\cite{FultonHarris91}, and $\chi^\mu(\pi)=1$ for all $\pi$. It follows
that 
\[
\chi^{\lambda \mu}(q)=(-1)^n\sum_\pi (-q)^{c(\pi)}=q(q-1)
\ldots (q-n+1).
\]
Thus $\chi^{\lambda \mu}(q)=0$ for $q=1, \ldots, n-1$. Hence $(\bar
\lambda, \bar \mu) \in \Gamma^{p,n}$. For $q \ge n$, a state with the
appropriate spectra for $\rho_{AB}$ and $\rho_A$ is
$\rho_{AB}=\frac{1}{n}\proj{0}_A \otimes \sum_{i=1}^n \proj{i}_B$.

Since $\chi^{\lambda \mu}(q)=\chi^{\mu \lambda}(q)$, if $\lambda=(n)$,
$\mu=(1^n)$ then $(\bar \lambda, \bar \mu) \in \Gamma_n$. A state with
the appropriate spectra is $\rho_{AB}=\proj{\psi_{AB}}$, where
$\psi_{AB}=\frac{1}{\sqrt{n}}\ket{11+ \dots +nn}_{AB}$. (Note that
this form of $\mu$ implies $p \ge n$.)
\end{example}

We can extend Proposition \ref{zeros} as follows
\begin{proposition}
For any $\lambda \in \Par{n}{pq}$, $\mu \in \Par{n}{p}$, there is a
positive integer $q_+$ and a negative integer $q_-$ such that
$\chi^{\lambda \mu}(q) \ne 0$ for $q \ge q_+$ and $q \le q_-$, and
$\chi^{\lambda \mu}(q)=0$ for $q_- < q < q_+$. 
\end{proposition}
\begin{proof}
Let $\lambda^\prime$ denote the diagram conjugate to $\lambda$,
obtained by interchanging rows and columns. Then
$\chi^{\lambda^\prime}(\pi)=(-)^{n+c(\pi)}\chi^{\lambda}(\pi)$, so
$\chi^{\lambda^\prime \mu}(q)=(-1)^n\chi^{\lambda \mu}(-q)$. It
follows that the negative range of integral roots has the same
properties as the positive range, and the result follows from
Proposition \ref{zeros}.
\end{proof}

\begin{example}
Table \ref{polynomials} gives some examples of $\chi^{\lambda \mu}(q)$
for $n=5$, illustrating the fact that the integral roots form a
sequence without a gap. Note that $\chi^{\lambda^\prime
\mu^\prime}(q)=\chi^{\lambda \mu}(q)$, since
$\chi^{\lambda^\prime}(\pi)=(-)^{n+c(\pi)}\chi^{\lambda}(\pi)$. To
illustrate the property $\chi^{\lambda^\prime
\mu}(q)=(-1)^n\chi^{\lambda \mu}(-q)$, for each $(\lambda,\mu)$,
either $(\lambda^\prime,\mu)$ or $(\lambda,\mu^\prime)$ is also given.

For each $\lambda$, $\mu$, states with $q=q_+$ and the appropriate
spectra are described in Example \ref{same} for the cases where
$\lambda=\mu$, and in Example \ref{extreme} for the case $(5),
(1^5)$. States for the other cases are easy to construct; eg for
$(4,1), (2,1^3)$, where $q_+=3$, we can take $p=4$ and
\beastar
\rho_{AB}&=&\frac{1}{5}\proj{11}_{AB}+\frac{4}{5}\proj{\psi_{AB}},\\
 \mbox{ where } \ket{\psi}_{AB}&=&\frac{1}{2}\ket{22+33}_{AB}+\frac{1}{\sqrt{2}}\ket{41}_{AB}.\\
\eeastar
\end{example}

%Proposition \ref{zeros} tells us that for $q \ge n$, $\tr_{\complex^q}
%\rho_\lambda$ has a non-vanishing coefficient for $\rho_\mu$ for every
%diagram $\mu$ in $Par(n,p)$. From Proposition \ref{dual-deFinetti} we
%know that, for $q \gg n(n-1)$, $\tr_{\complex^q} \rho_\lambda \approx
%\cI/p^n$.  Since $\cI=\sum_\mu e_\mu f_\mu \rho_\mu$, we can say the
%following: For $q \ge n$ we know that every $\rho_\mu$ occurs in the
%trace with nonzero coefficient, and for $q \gg n(n-1)$ we are assured
%that this coefficient closely approximates $e_\mu f_\mu/p^n$. This
%gives us some information about the convergence of $\tr_{\complex^q}
%\rho$ to the fully mixed state.

\begin{table}
\caption{Some examples of the polynomials $\chi^{\lambda \mu}(q)$ for $n=5$.
\label{polynomials}}
\begin{center}
\begin{tabular}{|c|c|c|}
\hline
$\lambda$, $\mu$ & $\chi^{\lambda \mu}(q)$ & integral roots\\
\hline
\hline
$(5), (5)$ ; $(1^5), (1^5)$ & $q^5+10q^4+35q^3+50q^2+24q$ & $-4,-3,-2,-1,0$\\
\hline
$(5), (4,1)$ ; $(1^5), (2,1^3)$ & $4q^5+20q^4+20q^3-20q^2-24q$ & $-3,-2,-1,0,1$ \\
\hline
$(4,1), (4,1)$ ;$(2,1^3), (2,1^3)$& $16q^5+40q^4+20q^3+20q^2+24q$ & $-2,-1,0$\\
\hline
$(4,1), (2,1^3)$& $16q^5-40q^4+20q^3-20q^2+24q$ & $0,1,2$\\
\hline
$(5), (2,1^3)$; $(1^5), (4,1)$ & $4q^5-20q^4+20q^3+20q^2-24q$ & $-1,0,1,2,3$\\
\hline
$(5), (1^5)$ & $q^5-10q^4+35q^3-50q^2+24q$ & $0,1,2,3,4$\\
\hline
\end{tabular}
\end{center}
\end{table}

Turning now to shifted Schur functions, Horn's conjecture -- now a
theorem \cite{Klyachko98} -- states that, given $\lambda, \mu,
\nu \in \Par{n}{d}$, $c^\lambda_{\mu \nu}>0$ if and only if there is
triple of $n \times n$ Hermitian matrices $A$, $B$ and $C$ with
eigenvalues $\lambda$, $\mu$ and $\nu$, respectively, such that
$A+B=C$. Thus, if we know that $\sum_\nu c_{\mu \nu}^\lambda f_\nu
>0$, we can infer that
\begin{proposition}[Shifted Schur function condition for Horn's conjecture] \label{horn}
Suppose $\lambda \in \Par{n}{d}$, $\mu \in \Par{k}{d}$. Then
$s^*_\mu(\lambda) > 0$ implies that there are $n \times n$ Hermitian
matrices $A$, $B$ and $C$ such that $A+B=C$ and the eigenvalues of $C$
are $\lambda_i$, and those of $A$ are $\mu_i$.
\end{proposition}
\begin{proof}
By Proposition \ref{fsum}, $s^*_\mu(\lambda) > 0$ implies there is
some $\nu \in \Par{n-k}{d}$ such that $c^\lambda_{\mu \nu} >0$, and by
the Horn-Klyachko theorem there are Hermitian matrices $A$, $B$, $C$
with eigenvalues $\lambda$, $\mu$, $\nu$, respectively, satisfying
$A+B=C$.
\end{proof}
Unlike Proposition \ref{gamma}, there is a simple criterion for the
conditions of Proposition \ref{horn} to hold, since $s^*_\mu(\lambda)
> 0$ if and only if $\mu \subset \lambda$. This follows immediately
from the fact that $f_\lambda s^*_\mu(\lambda)/(n \downharpoonright
k)=\sum_\nu c^\lambda_{\mu \nu} f_\nu= \dim \lambda/\mu$, where $\dim
\lambda/\mu$ is the number of standard numberings of the skew diagram
$\lambda/\mu$. This is a positive integer when $\mu \subset \lambda$
and zero otherwise. Indeed, when $\mu \subset \lambda$ it is clear
that the matrix $B$ with $\lambda_i-\mu_i$ down the diagonal satisfies
$A+B=C$, where $A$ and $C$ are diagonal with spectra $\mu$ and
$\lambda$, respectively. Thus this ``two matrix'' version of Horn's
conjecture of the single marginal problem is essentially trivial,
unlike its dual counterpart, the single marginal problem. However,
note that the single marginal problem is also trivial, by Proposition
\ref{zeros}, in the sense that the condition $\chi^{\lambda \mu}(q)>0$ is
always satisfied, unless one also specifies the dimension $q$ of the
traced-out subsystem.

%\begin{table}
%\caption{Convergence of coefficients of $\rho_\mu$ to $e^p_\mu
%f_\mu/p^n$ for $n=4$, $p=4$, $\lambda=(4)$ and various values of $q$.
%\label{converge}}
%\begin{center}
%\begin{tabular}{|c||c|c|c|c|c|}
%\hline
%q & $(1^4)$ & $(2,1^2)$ & $(2,2)$ & $(3,1)$ & $(4)$\\
%\hline
%\hline
%1 & 0 & 0 & 0 & 0 & 1\\
%\hline
%2 & 0 & 0 & 0.061 & 0.409 & 0.530\\
%\hline
%3 & 0 & 0.033 & 0.088 & 0.495 & 0.385 \\
%\hline
%4 & 0.0003 & 0.058 & 0.103 & 0.523 & 0.316\\
%\hline
%10 & 0.0017 & 0.120 & 0.134 & 0.542 & 0.203\\
%\hline
%100 & 0.0036 & 0.170 & 0.154 & 0.530 & 0.143\\
%\hline
%\hline
%$\infty$&0.0039 & 0.176 & 0.156 & 0.527 & 0.137\\
%\hline
%\end{tabular}
%\end{center}
%\end{table}

%
%\begin{example}
%When $\lambda=(n)$, $\chi^\lambda(\pi)=1$, for all $\pi$, so
%\[
%\chi_q^{\lambda \mu}=\sum_\pi 
%q^{c(\pi)}\chi^\lambda(\pi) \chi^\mu(\pi)=\sum_\pi
%q^{c(\pi)} \chi^\mu(\pi)=n!e^q_\mu,
%\]
%the last step by equation (\ref{e-formula}). Thus the coefficient of
%$\rho_\mu$ in $\tr_{\complex^q} \rho_\lambda$ is $e^p_\mu
%e^q_\mu/e^{pq}_\lambda$, which is given for $n=4$, $p=4$ and various
%values of $q$ in Table \ref{converge}.
%\end{example}

%
%
%
%                       GENERAL WERNER STATES
%
%
%
\section{General Werner states.} \label{general}

We now drop the assumption that the state is symmetrical, and consider
a general Werner state. First we characterise such states. 
%Let
%$U_\lambda \otimes V_\lambda$ be the $\lambda$-isotypic component in
%$(\complex^d)^{\otimes n}$. We can write this as
%$\bigoplus_{i=1}^{f_\lambda} U^i_\lambda$, where the $U^i_\lambda$ are
%copies of $U_\lambda$.
\begin{proposition} [Werner state characterisation]\label{Werner-form}
Any Werner state $\rho$ can be written
\[
\rho= \sum_\lambda \sum_i r_\lambda^i P^i_{U_\lambda},
\]
where $r_\lambda^i$ are positive constants, and $P^i_{U_\lambda}$ are
projectors onto unitary irreps $U^i_\lambda$.
\end{proposition}
\begin{proof}
If $\rho=\sum \gamma_i \proj{a_i}$ is the eigenvalue decomposition of
$\rho$, unitary invariance implies
\[
\rho=\sum \gamma_i \twirl(\proj{a_i}).
\]
From the Schur-Weyl
decomposition, we can write
\[
\ket{a_i}=\sum_{\lambda} \gamma_{i,\lambda} \ket{a_{i,\lambda}},
\]
and Schur's lemma then tells us that
\[
\twirl(\proj{a_i})=\sum_{\lambda} |\gamma_{i,\lambda}|^2 \twirl(\proj{a_{i,\lambda}}).
\]
Let $U^i_\lambda$ be the subspace generated by the set
$\{U\ket{a_{i,\lambda}}\ |\ U \in {\cal U}(d)\}$. This is a unitary
irrep, and $\twirl(\proj{a_{i,\lambda}})$ is an intertwining operator from
$U^i_\lambda$ to itself, and hence by Schur's lemma is proportional to
the projector $P^i_{U_\lambda}$.
\end{proof}
\begin{corollary} \label{Werner-dimension}
The number of (real) degrees of freedom of the set of Werner states
on $(\complex^d)^{\otimes n}$ is $d_W=\sum f_\lambda^2 -1$, where the
sum is over $\lambda$ in $\Par{n}{d}$.
\end{corollary}
\begin{proof}
Another way of stating the result of the Proposition is that the
$\lambda$-isotypic part of any Werner state is isomorphic to $\rho
\otimes P_{U_\lambda}$, where $\rho$ is any density matrix on
$V_\lambda$. But $\rho$ has $f_\lambda$ real terms down the diagonal,
with one constraint due to the sum of eigenvalues being 1, and there
are $f_\lambda(f_\lambda-1)$ real components in the non-diagonal terms
above the diagonal, and those below the diagonal are the conjugates of
those above.
\end{proof}
We are now ready to prove the main theorem:
\begin{theorem}[General dual de Finetti theorem] \label{general-dual}
If $\rho$ is a Werner state on $(\complex^p \otimes
\complex^q)^{\otimes n}$ and $q \ge n$, then
\[
||\tr_{\complex^q} \rho - \frac{\cI}{p^n}|| \le  2-2\left(\frac{q-n+1}{q}\right)^n = \frac{2n(n-1)}{q} + O(n^4/q^2).
\]
\end{theorem}
\begin{proof}
By Proposition \ref{Werner-form}, it suffices to consider a state $\rho$ that
is a normalised projector onto a unitary irrep, i.e. a state of the
form
\be \label{rho}
\rho=P_{U_\lambda}/e^{pq}_\lambda. 
\ee
Let $\{ a_i\}$ and $\{b_j\}$ be bases for $\complex^p$ and
$\complex^q$, respectively. We can define the Cartan subgroup of
$\cU(pq)$ as the set of matrices diagonal in the product basis $\{ a_i
\otimes b_j\}$. Let $\cF$ denote the set of lexicographically ordered
$n$-tuples of elements of this basis, which we write as $((i_1j_1)
\dots (i_nj_n))$; these define the weights of $U_\lambda$.
Let $\cD$ be the subset of $\cF$ where the $j$ indices are distinct;
this set is non-empty because we are assuming $q \ge n$. The
corresponding weight vectors are linear combinations of terms whose
indices are permutations of those that occur in the weight, i.e.
$((i_{\pi(1)}j_{\pi(1)}) \dots (i_{\pi(n)}j_{\pi(n)}))$ for some
permutation $\pi \in S_n$.

Let $U^\cD_\lambda$ be the subspace of $U_\lambda$ consisting of all
the weight spaces for elements of $\cD$. A permutation of the product
basis $\{ a_i \otimes b_j\}$, which can be regarded as an element of
$S_{pq}$, induces a unitary map on $U^\cD_\lambda$, and hence
$P_{U_\lambda}^\cD$, the projector on $U^\cD_\lambda$, is invariant
under such permutations. This implies that terms of the form
\[
\left(\proj{a_{i_{\pi(1)}}} \otimes \proj{b_{j_{\pi(1)}}}\right) \otimes \dots  \otimes \left(\proj{a_{i_{\pi(n)}}} \otimes \proj{b_{j_{\pi(n)}}}\right)       
\]
in $P_{U_\lambda}^\cD$ all have the same coefficients, since any two
such terms with different permuations $\pi$ in $S_n$ can be mapped
into each other by an appropriate basis permutation in $S_{pq}$. Thus
$tr_{\complex^q} P_{U_\lambda}^\cD$, i.e. the result of tracing out
the $\ket{b_j}s$ from $P_{U_\lambda}^\cD$, is a sum of terms
\[
\proj{a_{i_{\pi(1)}}}  \otimes \dots  \otimes \proj{a_{i_{\pi(n)}}} ,      
\]
for all $\pi \in S_n$, all terms having equal coefficients. Therefore
$tr_{\complex^q} P_{U_\lambda}^\cD$ is proportional to the identity
$\cI$ on $(\complex^p)^{\otimes n}$.

Now $U^\cD_\lambda$ is the union of weight spaces, all of which are
isomorphic and have dimension given by the Kostka number
$K_{\lambda,(1^n)}$, which is $f_\lambda$ (see
\cite[p. 56-57]{FultonHarris91}). As there are $p^n$ sets of possible
$i$-indices in $\cD$ and ${q \choose n}$ sets of distinct $j$-indices,
$U^\cD_\lambda$ has dimension $f_\lambda {q \choose n}p^n$. Thus,
\[
tr_{\complex^q} P_{U_\lambda}^\cD= f_\lambda {q \choose n}\cI.
\]
From this and eq. (\ref{rho}),
\[
tr_{\complex^q}\rho=\frac{tr_{\complex^q} P_{U_\lambda}}{e^{pq}_\lambda}=\frac{tr_{\complex^q} P_{U_\lambda}^\cD}{e^{pq}_\lambda} + A=\frac{f_\lambda {q \choose n}\cI}{e^{pq}_\lambda}+A,
\]
where A is a positive operator comes from tracing out the
remaining weight subspaces in $P_{U_\lambda}-P_{U_\lambda}^D$. Thus,
from the triangle inequality
\[
||tr_{\complex^q}\rho - \frac{\cI}{p^n}|| \le \left(1- \frac{f_\lambda {q \choose n}p^n}{e^{pq}_\lambda}\right)+||A|| = 2 \left(1- \frac{f_\lambda {q \choose n}p^n}{e^{pq}_\lambda}\right).
\]
The remainder of the proof consists in finding a lower bound for
$f_\lambda {q \choose n}/e^{pq}_\lambda$. To do this, we use the Weyl
dimension formula for $e^{pq}_\lambda$ and the hooklength formula for
$f_\lambda$ \cite{FultonHarris91} to write
\[
\frac{f_\lambda}{e^{pq}_\lambda}=\frac{n!(pq-1)!(pq-2)! \ldots
1!}{(pq+\lambda_1-1)!(pq+\lambda_2-2)! \ldots \lambda_d!}.
\]
This ratio decreases when a box in the diagram $\lambda$ is moved
upwards, so it achieves its minimum for the diagram $(n)$, giving
\[
\frac{f_\lambda}{e^{pq}_\lambda}
\ge \frac{n!(pq-1)!}{(pq+n-1)!}.
\]
Combining this with the inequality
\[
\frac{p(q-i+1)}{(pq+n-i)} \ge \frac{q-n+1}{q},
\]
which holds for $1 \le i \le n$, one concludes
\be \label{finalbound}
\frac{f_\lambda}{e^{pq}_\lambda} {q \choose n}p^n = \frac{q(q-1) \ldots (q-n+1) p^n}{(pq+n-1)(pq+n-2) \ldots pq} 
 \ge \left(\frac{q-n+1}{q}\right)^n.
\ee
\end{proof}

\begin{example} \label{tableau}

\begin{figure}
\centerline{\epsfig{file=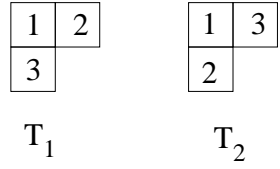,width=0.2\textwidth}}
\caption{The two tableaux for  $(2,1)$; see Example \ref{tableau}.
\label{young}}
\end{figure}

The simplest diagram $\lambda$ where $V_\lambda$ is non-trivial is
$(2,1)$. Here $f_\lambda=2$, corresponding to the fact that there are
two standard tableaux (numberings of $\lambda$ that increase downwards
and to the right), shown in Figure \ref{young} as $T_1$ and $T_2$. Let
$U_{\lambda,1}$ denote the unitary representation obtained by applying
the Young projector \cite{FultonHarris91} for the tableau $T_1$. The
normalised projector $\rho=P_{U_{\lambda,1}}/e^{pq}_\lambda$ onto this
representation is an example of a Werner state that is not
symmetric. We explicitly calculate an approximation to the trace
$tr_{\complex^q} \rho$ of this state.

As in the above proof, let $\{ a_i\}$ and $\{b_i\}$ be bases for
$\complex^p$ and $\complex^q$. Given $(i_1,i_2,i_3)$ and
distinct $(j_1,j_2,j_3)$, let us write
\[
\ket{u_{xyz}}=(\ket{a_{i_x}} \otimes \ket{b_{j_x}}) \otimes (\ket{a_{i_y}} \otimes \ket{b_{j_y}}) \otimes (\ket{a_{i_z}} \otimes \ket{b_{j_z}}),
\]
where $x,y,z$ is some permutation of $1,2,3$. Applying the Young
projector to the $\ket{u_{xyz}}$ for all possible permutations of
$1,2,3$ gives the set of vectors
\beastar
\ket{\psi_1}&=&\left(\ket{u_{123}}+\ket{u_{213}}-\ket{u_{321}}-\ket{u_{312}}\right)/2,\\
\ket{\psi_2}&=&\left(\ket{u_{132}}+\ket{u_{312}}-\ket{u_{231}}-\ket{u_{213}}\right)/2,\\
\ket{\psi_3}&=&\left(\ket{u_{321}}+\ket{u_{231}}-\ket{u_{123}}-\ket{u_{132}}\right)/2,\\
\eeastar 
These are linearly dependent, since
$\ket{\psi_1}+\ket{\psi_2}+\ket{\psi_3}=0$, and make the same angle
with each other, since $\braket{\psi_i}{\psi_j}=-1/2$ for all $i \ne
j$. Thus the projector onto the 2D subspace they span is
\[
\frac{2}{3}\left(\proj{\psi_1}+\proj{\psi_2}+\proj{\psi_3}\right).
\]
Summing this expression over all  $(i_1,i_2,i_3)$ and distinct
$(j_1,j_2,j_3)$ gives the projector $P_{U^D_{\lambda,1}}$.

Observe that $P_{U^D_{\lambda,1}}$ is not symmetric; for instance
$\ket{u_{123}}\bra{u_{321}}$ occurs with coefficient $-1/3$, whereas
$\ket{u_{213}}\bra{u_{231}}$ has coefficient $1/6$. However,
$\proj{u_{xyz}}$ has the same coefficient, $1/3$, for all permutations
$x,y,z$ of $1,2,3$, and it is only these terms that contribute to the
trace $tr_{\complex^q} P_{U^D_{\lambda,1}}$. Summing over
distinct indices $(j_1,j_2,j_3)$ we therefore find
\[
tr_{\complex^q} P_{U^D_{\lambda,1}}=3!{q \choose 3}\left(\frac{1}{3}\right) \cI.
\]
The factor $3!$ here arises because, for distinct $(i_1,i_2,i_3)$,
there are $3!$ ways of combining them with a set of distinct
$(j_1,j_2,j_3)$; eg $(i_1j_1,i_2j_2,i_3j_3)$,
$(i_1j_2,i_2j_1,i_3j_3)$, etc. When $(i_1,i_2,i_3)$ are not distinct,
there are fewer ways of combining them with $(j_1,j_2,j_3)$,
but on tracing out we regain the lost factor.

Since $e^d_\lambda=d(d-1)(d+1)/3$ for $\lambda=(2,1)$, we can write
\[
tr_{\complex^q}
\frac{P_{U^D_{\lambda,1}}}{e^{pq}_\lambda}=\left[\frac{(q-1)(q-2)}{(q-1/p)(q+1/p)}\right]\frac{\cI}{p^3}.
\]
Let $\alpha$ denote the term in square brackets. Then
\[
tr_{\complex^q} \rho = \alpha \frac{\cI}{p^3} + A,
\]
and we see that $\alpha \to 1$ for $q \to \infty$. 
\end{example}
To conclude this section, we look at the dual to Proposition
\ref{Werner-form} and its corollary.
\begin{proposition} [Symmetric state characterisation]\label{Symmetric-form}
Any symmetric state $\rho$ can be written
\[
\rho= \sum_\lambda \sum_i r_\lambda^i P^i_{V_\lambda},
\]
where $r_\lambda^i$ are positive constants, and $P^i_{V_\lambda}$ are
projectors onto irreps $V^i_\lambda$ of the symmetric group.
\end{proposition}
\begin{corollary} \label{symmetric-dimension}
The number of degrees of freedom of the set of symmetric states on
$(\complex^d)^{\otimes n}$ is $d_S=\sum
(e^d_\lambda)^2 -1$, where the sum is over $\lambda$ in $\Par{n}{d}$.
\end{corollary}
One might wonder if the standard deFinetti theorem could be proved by
methods like those used for Theorem \ref{general-dual}. It seems that
this is not possible, as the symmetric group representations have no
analogue of the weight spaces that are essential for this proof.

%
%
%
%                         DISCUSSION
%
%
%
\section{Discussion}

The de Finetti theorem and its dual seem very different in
character. In the case of a symmetric Werner state $\rho_\lambda$, the
standard de Finetti theorem tells us that $\tr_{n-k} \rho_\lambda$,
the residual state when $n-k$ subsystems are traced out, can be
approximated by the twirled power state $\twirl(\sigma^{\otimes k})$,
where $\sigma$ has spectrum $\bar \lambda$ (see Proposition
\ref{to-twirl}). If one carries out a measurement on
$\twirl(\sigma^{\otimes k})$ of the projections onto the subspaces
$U_\mu \otimes V_\mu$ in the Schur-Weyl decomposition of
$(\complex^d)^{\otimes k}$, the measured $\mu$, normalised to $\bar
\mu$, approximates $\bar \lambda$ \cite{KeyWer01PRA,Alicki88}. One
will only get an accurate estimate if $k \gg d$; when this condition
is satisfied, most of the information about the initial state is
encoded in the traced-out state. By contrast, when part of each
subsystem of a unitary-invariant state is traced out, the resulting
state approximates a fully mixed state, which conveys no information
about the initial state.

One might wonder whether this difference between the standard and dual
de Finetti theorems is related to the number of parameters, $d_S$ and
$d_W$, needed to specify symmetric and Werner states, respectively. Is
there a large reduction in $d_W$ in tracing out $\complex^q$ from each
subsystem? If so, the loss of information about the initial state
would be explained. However, this is not the case. In fact, for $p >
n$, $d_W$ is given by $\sum f_\lambda^2 -1$ over $\lambda \in
\Par{n}{d}$ (Corollary \ref{Werner-dimension}), and is the same for
the whole system, where $d=pq$, and for the traced-out system where
$d=p$. There is actually more of a reduction in the number of
parameters with the standard de Finetti theorem, since $d_S$ is given
by $\sum_{\lambda \in \Par{n}{d}} (e^d_\lambda)^2-1$ (Corollary
\ref{symmetric-dimension}), which does increase, though only
polynomially, with $n$.

For the approximation to the fully mixed state to be
close, the dimension $q$ of the traced-out part of each subsystem must
be large relative to $n(n-1)$, where $n$ is the number of
subsystems. Note that one does not require that $p/q$ is small, where
$p$ is the dimension of the remaining part of each subsystem after
tracing-out. The situation is therefore not directly analogous to the
standard de Finetti theorem, where a good approximation requires that
$(n-k)/n$, the ratio of the number of subsystems traced out to the
total number of subsystems, be close to 1.

When $n=1$, the bound in the dual de Finetti theorem is zero, which
tells us that no tracing-out is needed; this just conveys the familiar
fact that averaging the action of ${\cal U}(d)$ on a state on
$\complex^d$ gives the fully mixed state. One can ask which finite
subsets $S$ of ${\cal U}(d)$ have the property that the average
$\sum_S U \rho U^{\dagger}/|S|$ gives a good approximation to the
fully mixed state for any $\rho$, and it is known \cite{HLSW04} that
there are such sets with $|S| \approx d\log d$. The same question can
be posed for $n>1$, though now we expect to have to trace out part of
each subsystem to get an approximation to the completely mixed state.

The dual de Finetti theorem has a certain resemblance to a theorem
proved in \cite{PopescuShortWinter05}. This asserts that if $H_E$, the
state space of the environment, is traced-out from a random state
$\rho$ on the product of the system and environment $H_S \otimes H_E$,
then $\tr_E \rho$ is approximately a fully mixed state, the
approximation improving as $\dim H_E/\dim H_S$ increases. (Actually
the theorem holds more generally, for a state defined on an arbitrary
subspace of $H_S \otimes H_E$.) This suggests that obtaining the fully
mixed state after tracing out should be a property that holds for
``almost all states'', and not just for those with the special
structure of Werner states. One might therefore hope to be able to
extend the dual de Finetti theorem to a larger class of states (though
mathematics abounds with propositions known to be almost always true,
yet where specific instances are rather hard to find).

A natural application is to quantum secret-sharing: the theorem tells
us that this can be achieved by splitting up the subsystems of a Werner
state and giving them to two or more parties. With two parties, for
instance, each can have half of each subsystem, though the dimension
of each subsystem has to be large relative to $n$ for this to
work. Note that the procedure relies on the fact that $p/q$ does not
have to be small; we need to be able to regard both $\complex^{\otimes
q}$ and $\complex^{\otimes p}$ as the traced-out part (and similarly
for more than two parties).

Finally, one can ask whether the de Finetti theorem and its dual are
facets of some more all-embracing version of the theorem.

\section{Acknowledgements}
I thank Matthias Christandl, Robert K\"onig, Renato Renner and Graeme
Segal for helpful comments. This work was supported by the
European Union through the Integrated Project QAP (IST-3-015848),
SCALA (CT-015714), and SECOQC.

\end{document}